\documentclass[final,3p,times]{elsarticle}

\usepackage{amssymb}
\usepackage{amsmath}
\usepackage{placeins}

\journal{Journal of Subatomic Particles and Cosmology}

\begin{document}

\begin{frontmatter}



\title{Third-body-corrected Two-Pion Femtoscopic
Correlations in Au+Au Collisions at High Baryon Density}

\author[aaa]{Youquan Qi for the STAR Collaboration}
\affiliation[aaa]{organization={Key Laboratory of Quark \& Lepton Physics (MOE) and Institute of Particle Physics, Central China Normal University},
             addressline={152 Luoyu Road},
             city={Wuhan},
             postcode={430079},
             state={Hubei},
             country={China}}

\begin{abstract}
Identical charged pion femtoscopy provides a sensitive probe of the space-time structure of particle-emitting sources in heavy-ion collisions. At high baryon density, the differences between $\pi^+\pi^+$ and $\pi^-\pi^-$ Hanbury Brown--Twiss (HBT) radii may originate from both the third-body Coulomb effect and the initial-state isospin asymmetry. Since the latter is related to the equation of state (EoS) of asymmetric nuclear matter, separating these two effects is essential. In these proceedings, we present the measurements of identical charged pion correlations in Au+Au collisions at $\sqrt{s_{\mathrm{NN}}}=3.0$, 3.2, 3.5, 3.9, 4.5, and 7.7 GeV collected by the STAR experiment. A correction procedure is developed to remove the residual third-body Coulomb effect. After correction, the extracted HBT radii of $\pi^+\pi^+$ and $\pi^-\pi^-$ become consistent within uncertainties. The results indicate that the observed charge dependence is dominated by the third-body Coulomb effect, while any remaining contribution from the initial-state isospin asymmetry to the relative HBT radius difference, $(R_{\pi^-}-R_{\pi^+})/R_{\pi^+}$, is constrained to be less than 5\% at the 95\% confidence level.
\end{abstract}


\end{frontmatter}

\section{Introduction}

Two-pion femtoscopy is a powerful tool to investigate the space-time evolution of particle-emitting sources in relativistic heavy-ion collisions. The extracted HBT radii provide information on the source size, collective expansion, and freeze-out dynamics~\cite{Lisa2005,STAR_HBT_BES}.

At low collision energies, where the system reaches high baryon density, STAR measurements have observed differences between $\pi^+\pi^+$ and $\pi^-\pi^-$ correlation functions and HBT radii~\cite{QiQM2023}. Such charge dependence may originate from two sources. One is the Coulomb interaction between emitted pions and the net positive charge of the emitting source (third-body Coulomb effect), which modifies the measured pion momenta. The other is the initial-state isospin asymmetry. At high baryon density, pion production is strongly influenced by $\Delta$ resonance decays, and the neutron excess in Au nuclei may lead to different emission properties for positive and negative pions. Since the isospin dependence of dense nuclear matter is closely related to the EoS~\cite{Li2008}, separating these two effects is important.

In this work, we develop a correction procedure to remove the third-body Coulomb effect for identical charged pion femtoscopy measurement. The corrected correlation functions and HBT radii are used to constrain possible charge-dependent effects from the initial-state isospin asymmetry.

\section{Analysis}

\subsection{Construction of two-pion correlation functions}

The data sample consists of Au+Au collisions at $\sqrt{s_{\mathrm{NN}}}=3.0$, 3.2, 3.5, 3.9, 4.5, and 7.7 GeV collected by STAR. Identical charged pion pairs are selected to construct three-dimensional correlation functions,

\begin{equation}
C(\mathbf{q})=\frac{A(\mathbf{q})}{B(\mathbf{q})},
\end{equation}

\noindent where $A(\mathbf{q})$ and $B(\mathbf{q})$ represent same-event and mixed-event relative momentum distributions, respectively. Event mixing is performed within the same centrality and vertex position classes to ensure similar event characteristics and detector acceptance. The correlation functions are analyzed in the LCMS (Longitudinally Co-Moving System) using the Bertsch--Pratt coordinates $(q_{\mathrm{out}},q_{\mathrm{side}},q_{\mathrm{long}})$ and fitted with the Bowler--Sinyukov parametrization~\cite{Sinyukov1998},

\begin{equation}
C(q)=N[(1-\lambda)+\lambda K(q)(1+G(q))],
\end{equation}

\noindent where $N$ is the normalization factor, $\lambda$ is the correlation strength parameter that accounts for the reduction of the correlation intercept due to effects such as particle misidentification and long-lived resonance decays, and $K(q)$ accounts for the Coulomb interaction between the two identical pions~\cite{Lednicky1982} and

\begin{equation}
G(q)=
\exp[-R_{\mathrm{out}}^2q_{\mathrm{out}}^2
-R_{\mathrm{side}}^2q_{\mathrm{side}}^2
-R_{\mathrm{long}}^2q_{\mathrm{long}}^2
-2R_{\mathrm{out,long}}^2q_{\mathrm{out}}q_{\mathrm{long}}].
\end{equation}

\noindent Here, the ``out'' direction is defined along the pair transverse momentum, the ``side'' direction perpendicular to it in the transverse plane, and the ``long'' direction along the beam axis. The HBT radii $R_{\mathrm{out}}$, $R_{\mathrm{side}}$, and $R_{\mathrm{long}}$ characterize the homogeneity lengths of the emission source in the corresponding directions.

\subsection{Third-body Coulomb correction}

\begin{figure}[!htbp]
\centering
\includegraphics[width=0.4\linewidth]{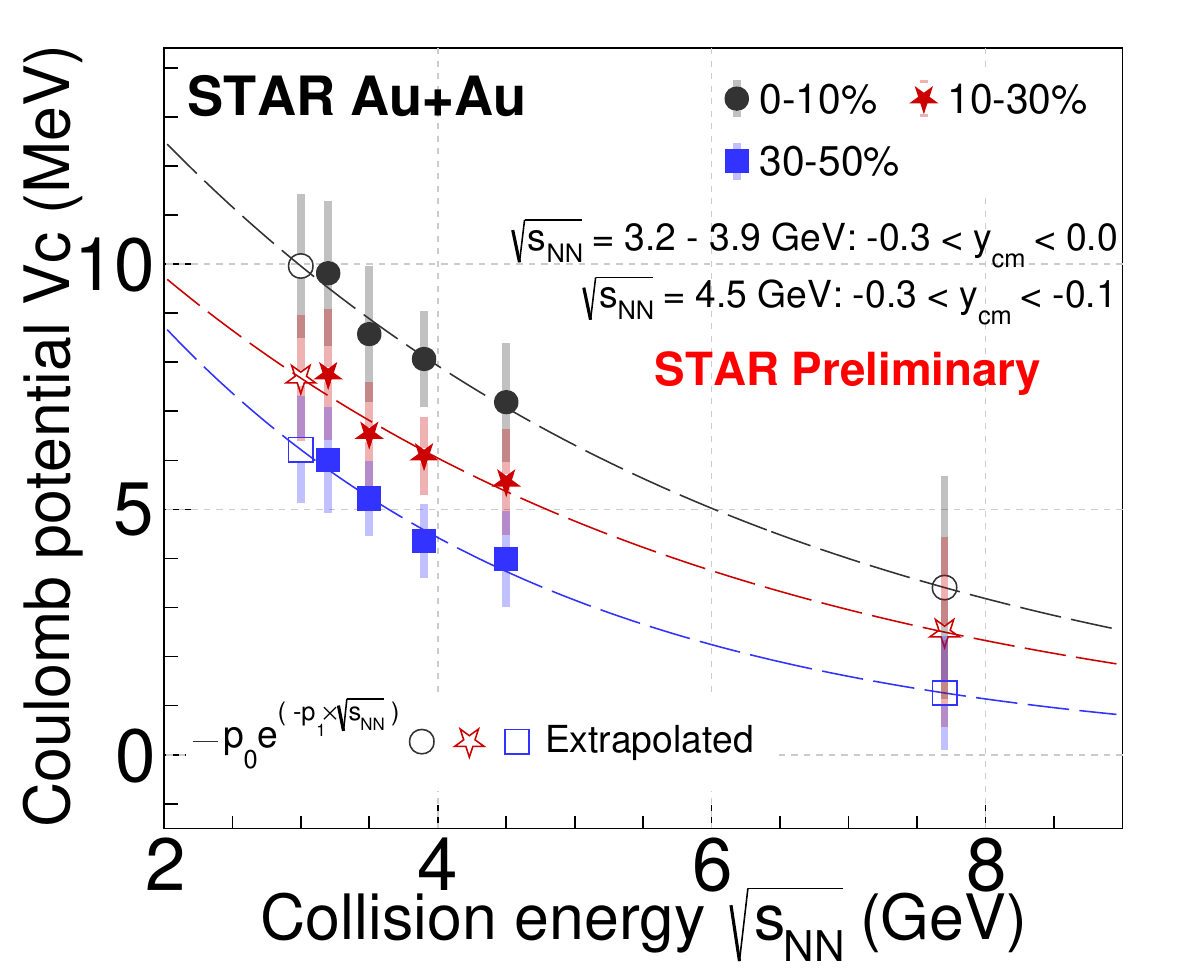}
\caption{
Collision energy dependence of the third-body Coulomb potential parameter $V_c$ used in the correction procedure. 
The black circles, red stars, and blue squares represent the results for 0--10\%, 10--30\%, and 30--50\% centrality, respectively. 
The solid markers indicate measured values, while the dashed curves show exponential-function fits. 
The open markers are obtained from extrapolation of the fitted functions. 
The shaded boxes represent the quadratic sum of statistical and systematic uncertainties.
}
\label{fig1}
\end{figure}

At low collision energies, baryon stopping leads to a positively charged source near mid-rapidity. The Coulomb interaction between this source and emitted pions modifies their momenta and introduces charge-dependent effects in the correlation functions. The momentum modification due to the third-body Coulomb interaction is evaluated using an effective Coulomb potential,
\begin{equation}
V_{\mathrm{eff}}
=
V_c
\left[
\mathrm{Erf}(x)-\frac{2}{\sqrt{\pi}}xe^{-x^2}
\right],
\end{equation}

\noindent where $x=\sqrt{\left(E_{\pi}/m_{\pi}-1\right)m_{p}/T_{p}}$, with $E_{\pi}$ and $m_{\pi}$ representing the pion energy and mass, respectively, $m_{p}$ denoting the proton mass, and $T_{p}$ representing the inverse-slope parameter extracted from the proton $m_{T}$ distribution~\cite{HADES:2022coulomb}. Due to baryon stopping, protons dominate the positively charged baryonic source and thus provide the main contribution to the third-body Coulomb field. The $V_c$ values used in this correction are extracted by simultaneously fitting the transverse momentum ($p_{\mathrm{T}}$) spectra of $\pi^-$ and $\pi^+$, taking into account the Coulomb interaction between the pions and the positively charged emitting source. The extracted $V_c$ values for different collision energies~\cite{ChenCPOD2026} are shown in Fig.~\ref{fig1}.
The corresponding momentum shift is calculated as

\begin{equation}
\Delta p=
\sqrt{(E+V_{\mathrm{eff}})^2-m^2}
-\sqrt{E^2-m^2}.
\end{equation}

\noindent The momentum shift is applied along the original pion momentum direction. The repulsive interaction between $\pi^+$ and the positively charged source increases the measured pion momentum, while the attractive interaction for $\pi^-$ decreases it. Therefore, to remove the third-body Coulomb effect, the momentum of $\pi^+$ is corrected by subtracting $\Delta p$, while the momentum of $\pi^-$ is corrected by adding $\Delta p$. The corrected pion momenta are then used to reconstruct the correlation functions and extract the HBT radii.

\section{Results and Discussion}

\begin{figure}[!htpb]
\centering
\includegraphics[width=0.45\linewidth]{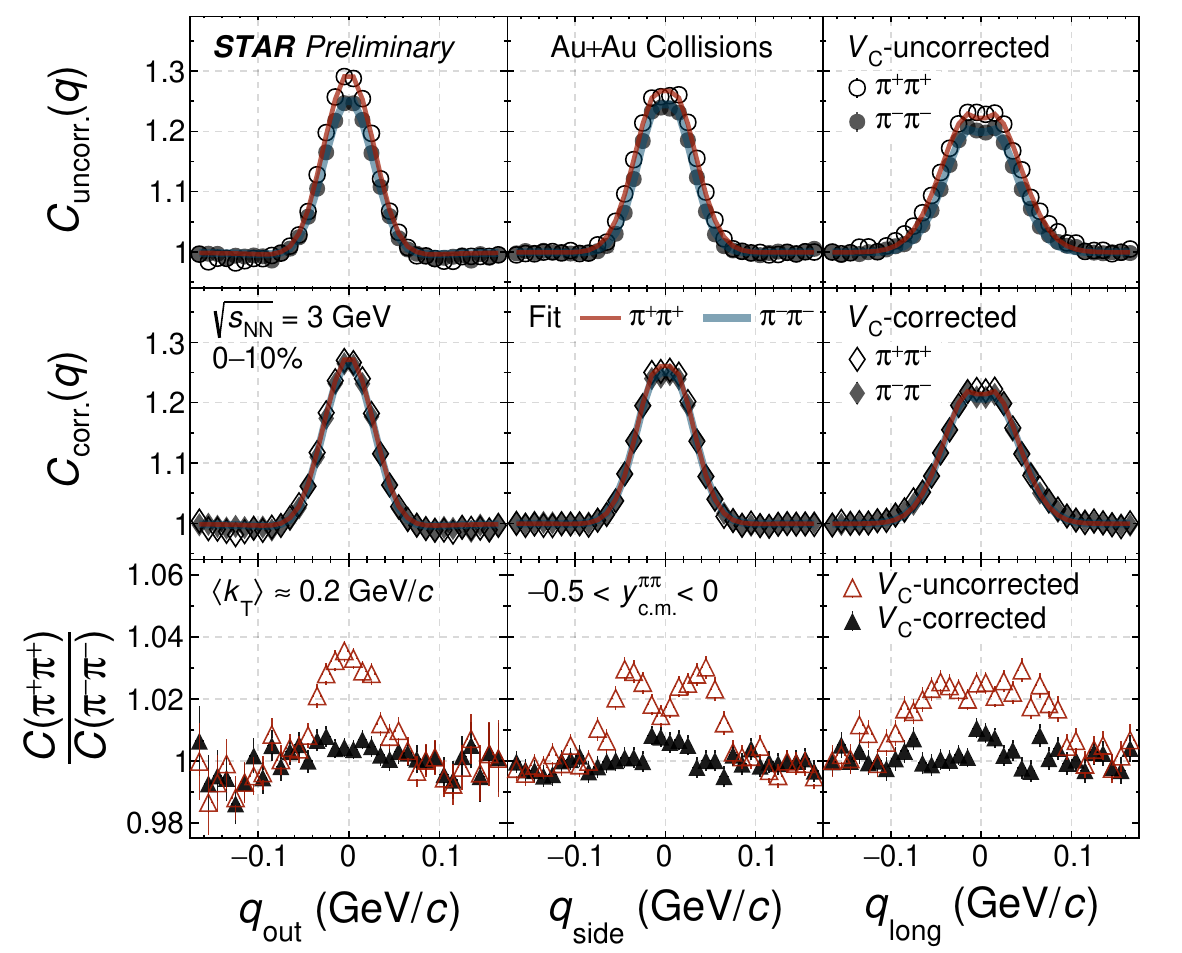}
\caption{
Two-pion correlation functions for $\pi^+\pi^+$ (open markers) and $\pi^-\pi^-$ (filled markers) pairs in Au+Au collisions at $\sqrt{s_{\mathrm{NN}}}=3.0$ GeV. 
The upper, middle, and lower panels show the $V_c$-uncorrected correlation functions, the $V_c$-corrected correlation functions, and the ratios of $\pi^-\pi^-$ to $\pi^+\pi^+$ correlation functions, respectively. 
The results are obtained for 0--10\% central collisions with $-0.5<y_{\mathrm{c.m.}}^{\pi\pi}<0$ and $0.15<k_{\mathrm{T}}<0.25$ GeV/$c$ as functions of $q_{\mathrm{out}}$, $q_{\mathrm{side}}$, and $q_{\mathrm{long}}$. 
The red and blue curves represent Gaussian fits to the $\pi^+\pi^+$ and $\pi^-\pi^-$ correlation functions, respectively.
}
\label{fig2}
\end{figure}

Figure~\ref{fig2} shows the $\pi^+\pi^+$ and $\pi^-\pi^-$ correlation functions before and after applying the third-body Coulomb correction in 0--10\% central Au+Au collisions at $\sqrt{s_{\mathrm{NN}}}=3.0$ GeV. Before the correction, clear differences between the two charge combinations are observed. The ratio of the $\pi^-\pi^-$ to $\pi^+\pi^+$ correlation functions deviates from unity, with a maximum difference of approximately 4\%. After applying the $V_c$ correction, the charge dependence is significantly reduced, and the correlation-function ratio closes to unity. This demonstrates that the observed charge splitting in the pion correlation functions is mainly caused by the third-body Coulomb effect.

\begin{figure}[!htbp]
\centering
\includegraphics[width=0.7\linewidth]{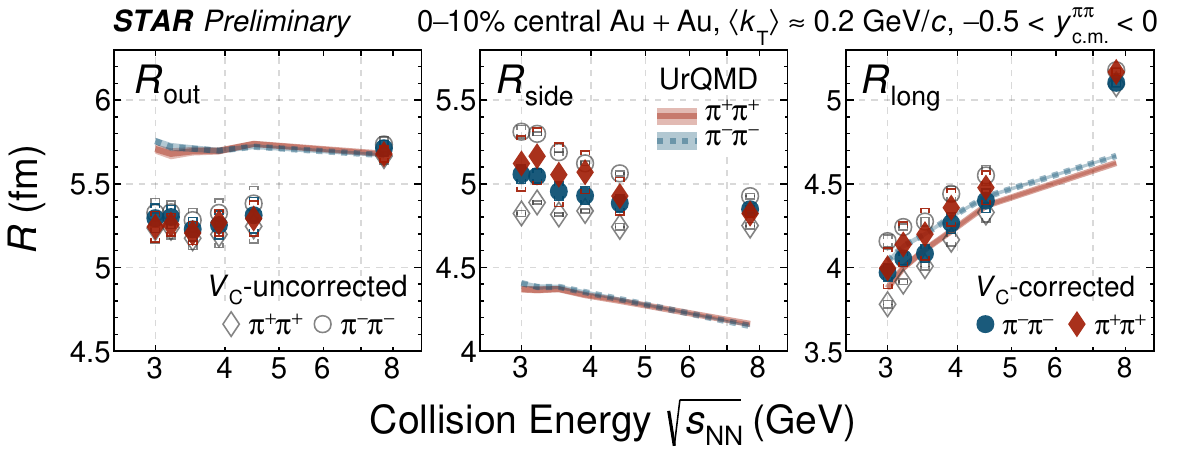}
\caption{
Collision energy dependence of $R_{\mathrm{out}}$, $R_{\mathrm{side}}$, and $R_{\mathrm{long}}$ in 0--10\% central Au+Au collisions for $\pi^+\pi^+$ (circles) and $\pi^-\pi^-$ (triangles) pairs. 
The $V_c$-corrected and uncorrected results are shown as open and filled markers, respectively. 
The statistical uncertainties are smaller than the marker size. 
The brackets and shaded boxes represent the systematic uncertainties of the corrected and uncorrected results, respectively. 
For comparison, UrQMD calculations for $\pi^+\pi^+$ and $\pi^-\pi^-$ pairs are shown as solid and dashed curves, respectively.
}
\label{fig3}
\end{figure}

Figure~\ref{fig3} presents the collision-energy dependence of the HBT radii extracted from $\pi^+\pi^+$ and $\pi^-\pi^-$ correlation functions. Before applying the correction, the radii, except $R_{\mathrm{out}}$, obtained from pairs $\pi^-\pi^-$ are systematically larger than those from pairs $\pi^+\pi^+$, especially at lower collision energies where the baryon density is higher. After the third-body Coulomb correction, the charge-dependent differences of $R_{\mathrm{side}}$ and $R_{\mathrm{long}}$ disappear within systematic uncertainties. This indicates that the observed difference between positive and negative pion femtoscopy is dominated by the third-body Coulomb effect. The residual contribution from the initial-state isospin asymmetry to the relative HBT radius difference, $(R_{\pi^-}-R_{\pi^+})/R_{\pi^+}$, is constrained to be less than 5\% at the 95\% confidence level for both $R_{\mathrm{side}}$ and $R_{\mathrm{long}}$.
The comparison between corrected and uncorrected results also shows that $R_{\mathrm{out}}$ is weakly affected by the third-body Coulomb correction, while the other radius components show a larger modification. The corrected $R_{\mathrm{long}}$ increases by approximately 20\% from $\sqrt{s_{\mathrm{NN}}}=3.0$ to 7.7 GeV, indicating an increase of the source lifetime and stronger longitudinal expansion with increasing collision energy. 

\section{Summary}

In summary, we have investigated the charge dependence of identical charged pion femtoscopy in Au+Au collisions at $\sqrt{s_{\mathrm{NN}}}=3.0$--7.7 GeV measured by the STAR experiment. A third-body Coulomb correction procedure was developed to remove the effect of the positively charged emitting source on pion momenta. Before the correction, significant differences between $\pi^+\pi^+$ and $\pi^-\pi^-$ correlation functions and extracted HBT radii were observed. After the correction, the charge dependent difference in the correlation functions is significantly reduced, and the extracted HBT radii agree within uncertainties. These results demonstrate that the observed charge splitting is dominated by the third-body Coulomb effect, while the possible contribution to the relative HBT radius difference, $(R_{\pi^-}-R_{\pi^+})/R_{\pi^+}$, from the initial-state isospin asymmetry is constrained to be less than 5\% at 95\% confidence level. The increase of $R_{\mathrm{long}}$ with collision energy suggests a longer source lifetime and enhanced longitudinal expansion at higher energies.
\section*{Acknowledgments}

This work is supported in part by the National Natural Science Foundation of China under Grant Nos. 12525509 and 12447102, the National Key Research and Development Program of China under Contract No. 2022YFA1604900, and the Fundamental Research Funds for the Central Universities (XJ2026000302).





\bibliographystyle{elsarticle-num}
\bibliography{sqm2026_template}



\end{document}